\documentclass{aa}
\usepackage{natbib}
\usepackage{graphicx}
\usepackage{amssymb,times,graphics, subfigure}
\usepackage[english]{babel}
\def\Rstar{\hbox{$R_{\star}$}}              

\bibpunct{(}{)}{;}{a}{}{,}

\begin{document}
   \title{Detection of `parent' molecules from the inner wind of AGB
   stars as tracers of non-equilibrium chemistry}


   \author{L.\ Decin\inst{1,2}\thanks{\emph{Postdoctoral Fellow of the
      Fund for Scientific Research, Flanders}} 
\and I.\ Cherchneff\inst{3} \and S.\ Hony\inst{4,1} \and
      S. Dehaes\inst{1}\thanks{\emph{Scientific Researcher of the Fund
      for Scientific Research, Flanders}} \and C.\ De Breuck\inst{5}
      \and K.~M.\ Menten\inst{6} }


   \institute{
  Department of Physics and Astronomy, Institute for Astronomy,
  K.U.Leuven, Celestijnenlaan 200B, B-3001 Leuven, Belgium\\
              \email{leen.decin@ster.kuleuven.be}
\and Sterrenkundig Instituut Anton Pannekoek, University of
  Amsterdam, Kruislaan 403 1098 Amsterdam, The Netherlands
\and Institut f\"ur Astronomie, ETH H\"onggerberg,
Wolfgang-Pauli-Strasse 16, 8093 Z\"urich, Switzerland\\
\email{isabelle.chechnneff@phys.ethz.ch}
\and Laboratoire AIM, CEA/DSM - CNRS - Université Paris Diderot,
DAPNIA/SAp, 91191 Gif sur Yvette, France \\
\email{sacha.hony@cea.fr}
\and European Southern Observatory, Karl-Schwarschild Strasse, D-85748
Garching bei M\"unchen, Germany\\
\email{cdebreuck@eso.org}
\and MPI f\"ur Radioastronomie, Auf dem H\"ugel 69, D-53121 Bonn,
Germany\\
\email{kmenten@mpifr-bonn.mpg.de}
              }

   \date{Received September 15, 2007; accepted October 15, 2007}

  \abstract
{Asymptotic Giant Branch (AGB) stars are typified by strong
  dust-driven, molecular outflows. For long, it was believed that the
  molecular setup of the circumstellar envelope of AGB stars is
  primarily determined by the atmospheric C/O ratio. However, recent
  observations of molecules such as HCN, SiO, and SO reveal gas-phase
  abundances higher than predicted by thermodynamic equilibrium (TE)
  models. UV-photon initiated dissociation in the outer envelope or
  non-equilibrium formation by the effect of shocks in the inner
  envelope may be the origin of the anomolous abundances.}
  {We aim at detecting \emph{(i)} a group of `parent' molecules (CO,
  SiO, HCN, CS), predicted by the non-equilibrium study of Cherchneff
  (2006) to form with almost constant abundances independent of the C/O
  ratio and the stellar evolutionary stage on the Asymptotic Giant
  Branch (AGB), and \emph{(ii)} few molecules, such as SiS and SO,
  which are sensitive to the O- or C-rich nature of the star.}
  {Several low and high excitation rotational transitions of key
  molecules are observed at mm and sub-mm wavelengths with JCMT and
  APEX in four AGB stars: the oxygen-rich Mira \object{WX Psc}, the S
  star \object{W Aql}, and the two carbon stars \object{V Cyg} and
  \object{II~Lup}. A critical density analysis is performed to
  determine the formation region of the high-excitation molecular
  lines.}
{We detect the four `parent' molecules in all four objects, implying
that, indeed, these chemical species form whatever 
the stage of evolution on the AGB. High-excitation lines of SiS are
also detected in three stars with APEX, whereas SO is only detected in
the oxygen-rich star \object{WX Psc}. }
{ This is the first multi-molecular observational proof that
  periodically shocked layers above the photosphere of AGB stars show
  some chemical homogeneity, whatever the photospheric C/O ratio and
  stage of evolution of the star. }

   \keywords{Astrochemistry, Molecular processes, Stars: AGB and post-AGB,
  (Stars): circumstellar matter, Stars: mass loss, Submillimeter}
\authorrunning{L.\ Decin et al.}
\titlerunning{Non-equilibrium chemistry in AGB stellar winds}

   \maketitle
%

\section{Introduction} \label{Introduction}
Circumstellar envelopes of Asymptotic Giant Branch stars (AGBs) have
 long been known to be efficient sites of molecule formation.  While
 the outer layers of such envelopes experience penetration of
 interstellar UV photons and cosmic rays resulting in a fast
 ion-molecule chemistry, the deepest layers are dominated by a
 non-equilibrium chemistry due to the passage of shocks generated by
 stellar pulsation. Dust forms in those inner gas layers, still bound
 to the star, and grains couple to the gas to accelerate it, thereby
 generating stellar wind and mass loss phenomena. The described
 processes greatly modify the abundances established by the
 equilibrium chemistry in the dense, hot photosphere
 \citep{Tsuji1973A&A....23..411T}.  

For a long time, the gas chemical composition was believed to be dominated
entirely by the C/O ratio of the photosphere. A C/O ratio greater than
one implied that all the oxygen was tied in CO, leading to an
oxygen-free chemistry, whereas a C/O ratio less than one meant that no
carbon bearing molecules apart from CO could ever form in an
oxygen-rich (O-rich) environment. This picture, based essentially on
thermal equilibrium considerations applied to the gas, has been first
disproved by the detection of SiO at millimeter (mm) wavelength in
carbon-rich (C-rich) AGBs \citep{Bujarrabal1994A&A...285..247B}. As
for O-rich AGBs, CO$_2$ infrared (IR) transition lines were detected
in various objects with the Short-Wavelength Spectrometer (SWS)
onboard the Infrared Space Observatory (ISO)
\citep[e.g.,][]{Justtaonont1996A&A...315L.217J, ryde1998Ap&SS.255..301R}.
   
Theoretical modeling describing the chemistry in the inner wind of the
extreme carbon star \object{IRC+10216} showed that the formation of SiO was
due primarily to hydroxyl OH reaction with atomic silicon close to the
photosphere as a result of shock activity and therefore
non-equilibrium chemistry \citep{Willacy1998A&A...330..676W}.  Later
on, \citet{Duari1999A&A...341L..47D} showed that CO$_2$ formation in
the O-rich Mira \object{IK Tau} results from the reaction of OH
radicals with CO in the shocked regions, implying again that
non-equilibirum chemistry was paramount to the formation of C-bearing
species in O-rich Miras. It was then recently proposed that the inner
wind of AGBs shows a striking homogeneity in chemical composition,
despite their photospheric C/O ratio and stage of stellar evolution
\citep{Cherchneff2006A&A...456.1001C}. In particular,
\citet{Cherchneff2006A&A...456.1001C} showed that when taking shock
chemistry into account, molecules such as SiO, HCN and CS are present
in comparable amount in the inner layers of M, S, and C AGBs, whereas
specific molecules (e.g.\ SO and HS for O-rich Miras and C$_2$H$_2$
for carbon stars) are typical for O-rich or C-rich chemistries.

In this letter, we present observations carried out with the JCMT and
the APEX telescope of four AGBs: one O-rich, \object{WX Psc}, one S
star ($\equiv$ C/O $\approx$ 1), \object{W Aql}, and two carbon stars,
\object{II Lup} and \object{V Cyg}. We focus on the detection of
(sub)mm transitions of CO, SiO, HCN, CS, SiS and SO in order to
confirm or disprove the above hypothesis and to check for homogeneity
in AGB winds.

\section{Observations and line profiles} \label{Observations}

The observations were performed in October 2006 with the 15\,m
JCMT\footnote{The James Clerk Maxwell Telescope (JCMT) is operated by
The Joint Astronomy Centre on behalf of the Science and Technology
Facilities Council of the United Kingdom, the Netherlands Organisation
for Scientific Research, and the National Research Council of
Canada. Program ID is m06bn03 (34\,h).} for \object{V Cyg}, \object
{WX Psc}, and \object{W Aql}, and in the period from September till
October 2006 with the APEX\footnote{APEX, the Atacama Pathfinder
Experiment, is a collaboration between the Max-Planck-Institut fur
Radioastronomie, the European Southern Observatory, and the Onsala
Space Observatory. Program ID is ESO.078.D-0534 (44\,h).} 12\,m
telescope for \object{II Lup}, \object{WX Psc} and \object{W Aql}. Due
to technical problems with the RxB3 JCMT receiver, only low frequency
lines within the RxA3 receiver (211 -- 276\,GHz) were obtained. For
the APEX observations, both the APEX-2A receiver (279--381\,GHz) and
FLASH receiver (460--495\,GHz and 780--887\,GHz) were used.  The
observations were carried out using a position-switching mode. The
JCMT data reduction was performed with the SPLAT devoted routines of
STARLINK, the APEX-data with CLASS.  A polynomial was fitted to an
emision free region of the spectral baseline and subtracted.  The
velocity resolution for the JCMT-data equals to 0.0305\,MHz, for the
APEX to 0.1221\,MHz. For \object{WX Psc}, \object{W Aql}, and
\object{II Lup} the data were rebinned to a resolution of 1\,km/s, for
\object{V Cyg} to 0.75\,km/s in order to have at least 40 independent
resolution elements per line profile. The antenna temperature,
$T_A^*$, was converted to the main-beam temperature ($T_{mb} =
T_A^*/\eta_{mb}$), using a main-beam efficiency $\eta_{mb}$ of 0.69
for the JCMT RxA3 receiver, of 0.73 for the APEX-2A receiver, and of
0.60 and 0.43 for the 460--495\,GHz and 780--887\,GHz FLASH channels
respectively \citep{Gusten2006A&A...454L..13G}.

\begin{figure*}
\centering
\includegraphics[width=.95\textwidth]{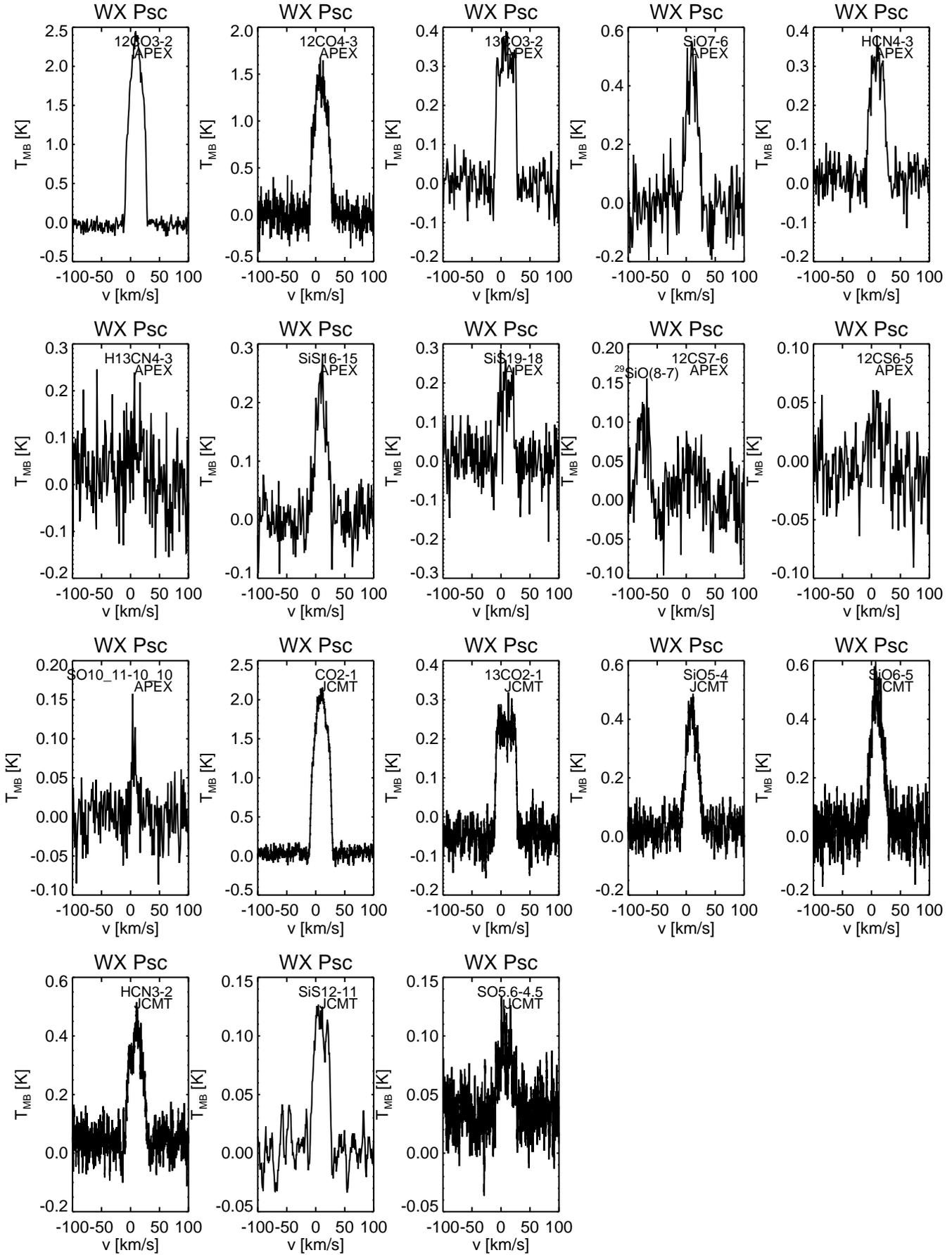}
 \caption{Molecular emission detected with APEX and JCMT in the O-rich
 AGB \object{WX Psc}.}  
\label{fig_WXPsc_all}
\end{figure*}

\begin{figure*}
\centering
\includegraphics{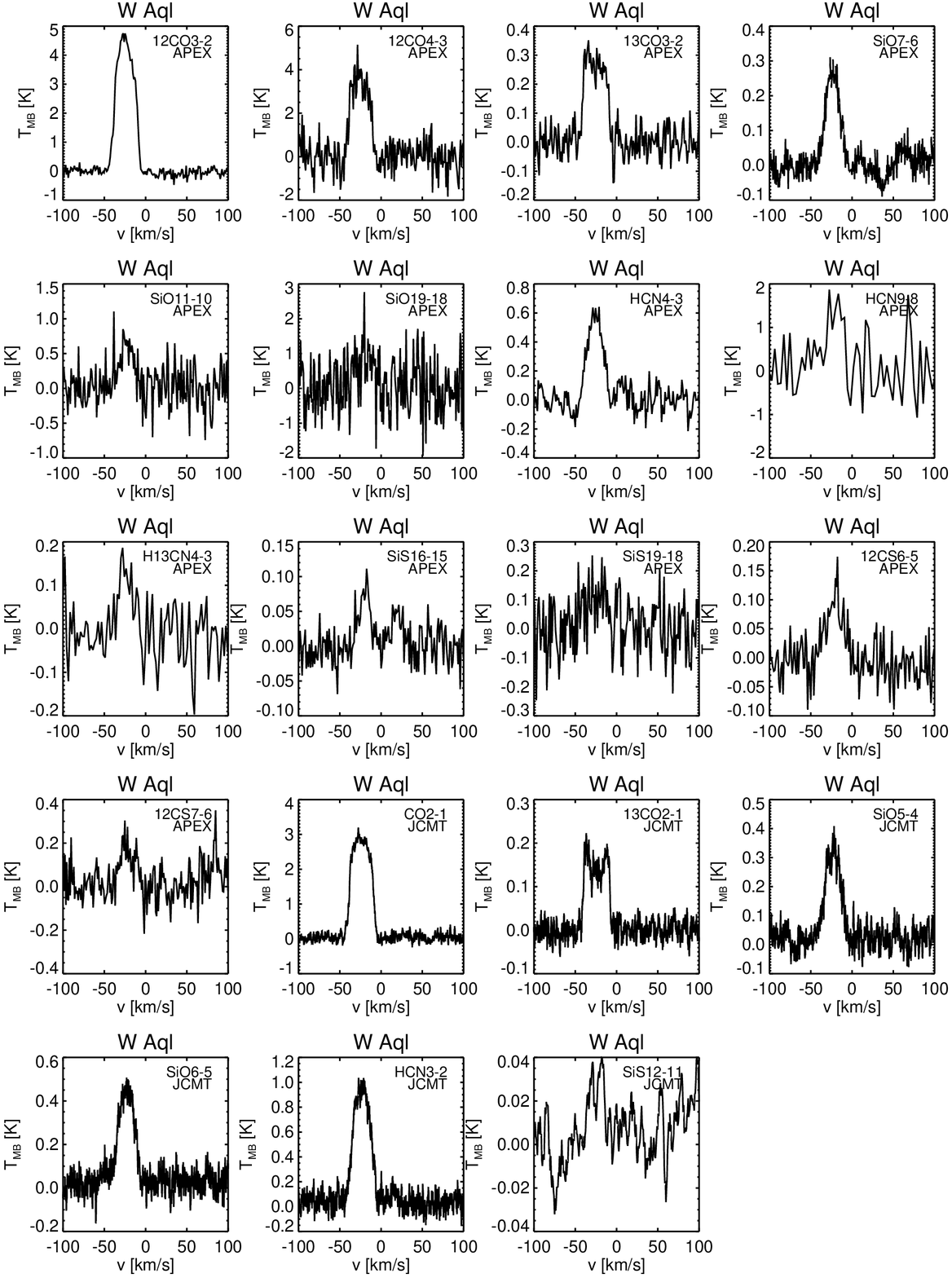}
 \caption{Molecular emission detected with APEX and JCMT in the S-type
 AGB \object{W Aql}. Note that the HCN(9-8) line is rebinned to a
 resolution of 3.6\,km/s, other lines are rebinned to a resolution of
 1.8\,km/s.}
\label{fig_WAql_all}
\end{figure*}

\begin{figure*}
\centering
\includegraphics[width=\textwidth,angle=90,height=.45\textheight]{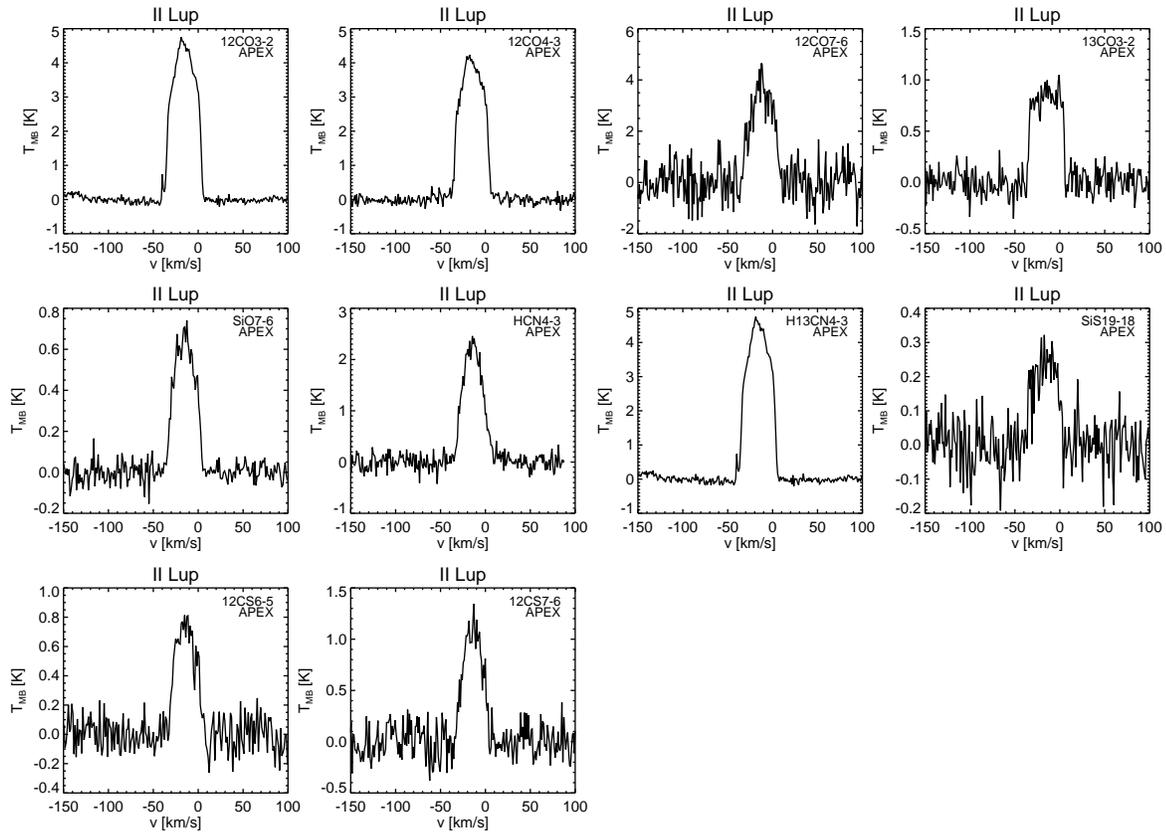}
 \caption{Molecular emission detected with APEX in the C-rich
 AGB \object{II Lup}.}  
\label{fig_IILup_all}
\end{figure*}

\begin{figure*}
\centering
\includegraphics[width=\textwidth,angle=90,height=.45\textheight]{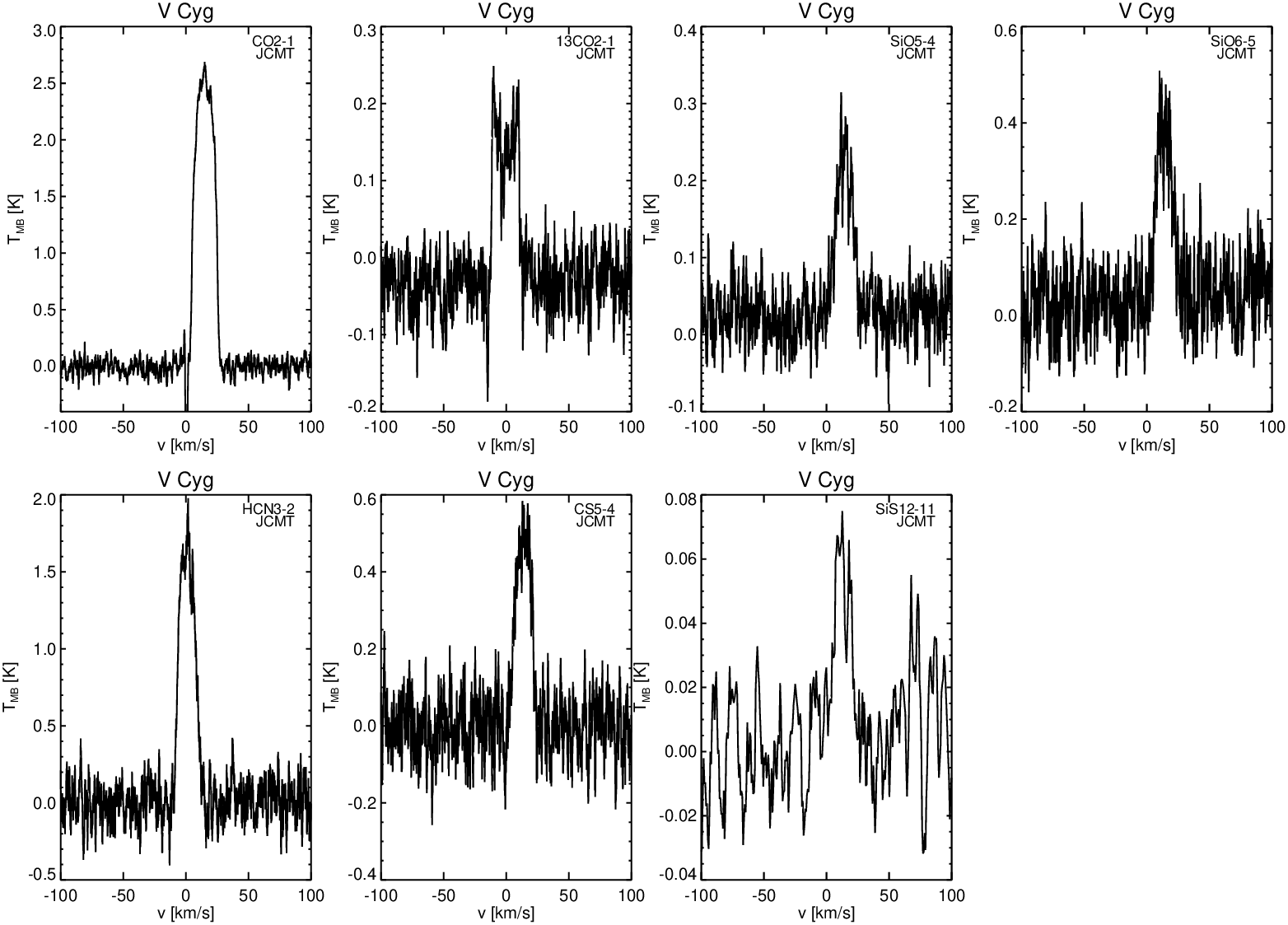}
 \caption{Molecular emission detected with JCMT in the C-rich
 AGB \object{V Cyg}.} 
\label{fig_VCyg_all}
\end{figure*}

The observed molecular emission lines of four AGB stars in our sample are 
displayed in the Figs.~\ref{fig_WXPsc_all} -- \ref{fig_VCyg_all}.

In all stars molecular emission lines of CO, SiO, HCN, and CS are
detected. This confirms the prediction of homogeneity by
\citet{Cherchneff2006A&A...456.1001C} as these species being `parent'
molecules which form in the inner layers of the CSE. This can be
understood in terms of the chemistry of these four molecules being
determined by shock propagation and not by the photospheric C/O ratio
and the stellar evolutionary stage.

SiS is also detected in all stars and we were able to detect with APEX
the high-excitation SiS(19-18) line in the O-rich \object{WX Psc}, the
C-rich \object{II Lup} and the S-rich \object{W Aql}. This is in good
agreement with recent SiS OSO, JCMT, and APEX observations of
\citet{Schoier2007arXiv0707.0944S} in a large sample of M and C stars,
including \object{WX Psc} and \object{V Cyg}. This implies that SiS
also forms close to the star, whatever the stage of stellar evolution.

Both the SO($6_5$ -- $5_4$) and the high-excitation
SO($10_{11}-10_{10}$) line were detected in O-rich \object{WX
Psc}. SO was neither found in the S-type \object{W Aql}, nor in the
two carbon stars \object{II Lup} and \object{V Cyg}. SO appears to be
typical for O-rich AGBs only, supporting the non-detection of SO in
C-stars by \citet{Woods2003A&A...402..617W}.

Both optically thin and optically thick lines occur (e.g.,
$^{13}$CO(2-1) versus the $^{12}$CO(2-1) line in \object{W Aql}, see e.g.\
Fig.~\ref{fig_WAql_all}).  The line parameters, i.e., the main-beam
brightness temperature at the line centre ($T_{mb}$), the line centre
velocity ($v_*$), and half the full line width ($v_e$), are obtained
by fitting the `soft parabola' line profile function to the data
\citep{Olofsson1993ApJS...87..267O}
\begin{equation}
T(v) = T_{\mathrm{mb}} \left[ 1 - \left( \frac{v-v_*}{v_{\mathrm e}} \right) ^2
\right] ^{\beta/2}\,,
\end{equation}
where $\beta$ describes the shape of the line. The velocity-integrated
intensities are obtained by integrating the emission between $v_* \pm
v_e$ and are listed in Table~\ref{Table_overview}. A value of
$\beta\,=\,2$ represents a parabolic line shape, expected for
optically thick expanding spherical envelopes in case the source is
much smaller than the radio telescope beam size; $\beta < 0$ results
in a profile with horns at the extreme velocity, expected for
optically thin lines when the source is resolved
\citep{Morris1975ApJ...197..603M}.  In our sample, several lines have
a $\beta$-value distinctly larger than 2, indicating that the line
formation occurs partially in the inner region, where the stellar wind
has not yet reached its full terminal velocity. This is notably the
case for the CS, SiO, and HCN lines in our sample, corroborating their
formation close to the star. The estimated terminal velocities for
\object{WX Psc}, \object{W Aql}, \object{II Lup}, and \object{V Cyg}
are 19.2\,km/s, 18.8\,km/s, 20.1\,km/s, and 12.4\,km/s, respectively.
A maximum of 5\,\% difference is found in the derived terminal
velocity values for each target, being within the velocity binsize,
indicating that all of the detected lines at least partly survive the
dust formation and wind acceleration processes. An exception is
SiS(19-18) in \object{WX Psc} (16\,km/s), but particularly the
SO($10_{11}$-$10_{10}$), with an estimated terminal velocity of only
10\,km/s, traces a much smaller geometrical region.

\begin{table*}
\caption{Overview of the velocity-integrated intensities ($\int
  T_{mb}\,dv$ in K\, km/s) for the observed line transitions. The
  frequency is listed in GHz and the lower energy level in
  cm$^{-1}$. The integrated intensity of detected lines with a low
  S/N-ratio is given between brackets. In case of a non-detection, an
  upper limit on the integrated intensity is computed as $3\sigma
  \times$ expected linewidth, with $\sigma$ the noise on the data.  A
  `$-$' indicates lines that were not observed. APEX-data are reported
  upright, JCMT-data are given in italics.}\label{Table_overview}
  \centering 
\setlength{\tabcolsep}{.3mm}
\vspace*{-2ex}
\begin{tabular}{ccccccccccccc }
\hline \hline
\emph{transition} & \emph{$^{12}$CO(2-1)} & $^{12}$CO(3-2) & $^{12}$CO(4-3) &
$^{12}$CO(7-6) & \emph{$^{13}$CO(2-1)} & $^{13}$CO(3-2) & \emph{$^{12}$CS(5-4)} &
$^{12}$CS(6-5) & $^{12}$CS(7-6) & $^{12}$CS(10-9) & $^{12}$CS(17-16)\\ 
\emph{frequency} & \emph{230.538} & 345.795 & 461.040  &  806.651 & \emph{220.398} & 330.587 & \emph{244.935} & 293.912 & 342.882 & 489.759 & 832.061\\
$E_{\rm{low}}$ & \emph{3.84} & 11.54 & 23.07 & 80.74 & \emph{3.67} &
11.03 & \emph{16.34} & 24.51 & 34.32  & 75.53 & 222.15 \\
\hline
WX Psc & \emph{58.0} & 64.3 & 40.2 & 22.4 & \emph{9.9} & 11.3 &\emph{ $<$2.4} & [0.8] & 1.3 & $<$8.1 & $<$42.6\\
W Aql &  \emph{80.6} & 118.4 & 95.9 & $<$178 & \emph{4.6} & 8.1 & $-$  & 2.4 & 4.6 & $<$54.1 & $<$203 \\
II Lup & $-$ & 145.0 & 130.0 & 99.7 & $-$ & 32.5 & $-$ & 20.4 & 35.2 & $-$ & $-$ \\
V Cyg & \emph{45.6} & $-$ & $-$ & $-$ & \emph{4.2} & $-$ &  \emph{7.4} & $-$ & $-$ & $-$ & $-$\\
\hline
\hline
\emph{transition}  & $^{13}$CS(7-6) &
$^{13}$CS(10-9) & \emph{H$^{12}$CN(3-2)} & H$^{12}$CN(4-3) & H$^{12}$CN(9-8)
& H$^{13}$CN(4-3) & \emph{SiO(5-4)} & \emph{SiO(6-5)} & SiO(7-6)  & SiO(11-10)& \\
\emph{frequency}   & 323.684 & 462.334 & \emph{265.886} & 354.505 &
797.433 & 345.340 & \emph{217.105} & \emph{260.518} & 303.926 & 477.504 & \\
$E_{\rm{low}}$ & 32.39 &  69.41 & \emph{8.87} & 17.74 & 106.42 &
17.28 & \emph{14.48} & \emph{21.73} & 30.42  & 79.65 & \\
\hline
WX Psc  & $<$67.4 & $<$12.7 & \emph{9.9}  & 8.9 & $<$117 & [2.6] & \emph{9.2} & \emph{11.1} & 10.4  & $<$15.0 & \\
W Aql  & $-$ & $<$13.0 & \emph{21.7} & 12.2 & [34.1] & 3.1 & \emph{7.1} & \emph{8.9} & 5.2 & 13.8 & \\
II Lup & $-$ & $-$ & $-$ & 60.7 & $-$ & 64.2 & $-$ & $-$ & 19.6  & $-$ & \\
V Cyg  & $-$ & $-$ & \emph{25.7} & $-$ & $-$ & $-$ & \emph{3.8} & \emph{5.1} & $-$ & $-$ & \\
\hline
\hline
\emph{transition}  &
SiO(19-18) & SiS(12-11) & \emph{SiS(14-13)} & SiS(16-15) & SiS(19-18) & SO($1_1$-$1_0$) &  \emph{SO($\mathit{6_5}$-$\mathit{5_4}$)} &
SO($7_8$-$6_7$) &  \emph{SO($\mathit{8_9}$-$\mathit{8_8}$)} & SO($10_{11}$-$10_{10}$) &  \\
\emph{frequency}   & 824.235 &
217.817 & \emph{254.102} & 290.380 & 344.778 & 286.34 &  \emph{219.949} & 340.714 &  \emph{254.573} & 336.597 & \\
$E_{\rm{low}}$   & 247.57 & 39.96 &
\emph{55.10} & 72.66 & 103.53 & 1.00 &  \emph{16.98} & 45.10 &  \emph{60.80} & 88.08& \\
\hline
WX Psc & $<$92.2 & \emph{3.2} & \emph{$<$1.2} & 5.4 & 4.8 & $<$0.9 &  \emph{1.6}
& $<$1.4 & $<$2.0 & [0.9] & \\
W Aql  & [22.3] & $-$ & $-$ & 1.2 & [3.4] & $-$ &\emph{$<$1.2} & $<$3.4 & $-$ & $-$ & \\
II Lup  & $-$ & $-$ & $-$  & $-$ & 7.8 & $-$ & $-$ & $<$4.2 & $-$ & $-$ & \\
V Cyg  & $-$ & \emph{0.9} & \emph{$<$14.8} & $-$ & $-$ & $-$ & \emph{$<$1.3} & $-$ & $-$ & $-$ & \\
\hline
\hline
\end{tabular}
\end{table*}


\section{Excitation analysis} \label{Analysis}

In case many transitions of an individual molecule can be observed, it
is possible to assess whether collisional or radiative excitation
mechanisms can produce the observed line intensities. While it is well
known that CO is formed in both M, C and S-type stars and survives dust condensation, it is of interest to study the excitation
requirements for the three other `parent' molecules SiO, HCN and CS
predicted by \citet{Cherchneff2006A&A...456.1001C} to be abundant in
the inner winds of M, S and C AGBs.

For all lines, except for those from the CO molecule, the emission
distribution is expected to be much smaller than the FWHM beam size of
the used telescope. To calculate column densities we need to correct
for the different beam filling factor, $f = \theta_S^2/(\theta_S^2 +
\theta_B^2)$, with $\theta_B$ and $\theta_S$ being the FWHM of the
beam and the source respectively. To do this, we define a
beam-averaged brightness temperature, $T_b$, scaled by correcting our
main-beam brightness temperatures for a fictitious $10''$ FWHM source,
i.e, $T_b = 1/f \times T_{mb}$.

Little interferometric data exist for the molecules that we have
observed in \textit{any} circumstellar envelope. We note, however,
that $4''$--$6''$ resolution observations of the HCN $J =1-0$ line in
the Mira variables TX Cam and IK Tau \citep{Marvel2005AJ....130..261M}
barely resolve the emission distibutions in these objects. Since,
first, those objects are closer to the Sun than our target stars and,
second, our higher exciation lines most likely arise from more compact
regions than the $1 - 0$ line, the column densities derived are strict
lower limits.

Assuming that the lines are optically thin and that the excitation
temperature, $T_{ex}$, between upper and lower level is such that
$T_{ex}$$\gg$$T_{bg}$ (with $T_{bg}$ the temperature of any background
source, e.g.\ 2.7\,K), the integration of the standard radiative
transfer equation shows that \citep{Goldsmith1999ApJ...517..209G}
\begin{equation}
N_u = \frac{1.67 \times 10^{14}}{\nu \ \mu^2} 
\left(\int{T_{b}\,dv}\right)\,,
\label{EqNup}
\end{equation} 
with $N_u$ the column density of the upper transition state in
cm$^{-2}$, $\mu$ the transition dipole moment in Debye, $\nu$ the transition
frequency in GHz, $T_{mb}$ the main-beam temperature in Kelvin and $v$
the velocity in km/s. For rotational transitions, the transition
moment for diatomic and polyatomic molecule is given by
\begin{equation}
\mu^2(J+1,J) = \frac{J+1}{2J+3}\, \mu_e^2\,,
\end{equation}
where $\mu_e$ is the permanent dipole moment of the molecule. Using the velocity--integrated intensities of Table~\ref{Table_overview}, we calculate the
resulting upper state column densities (see Table~\ref{Table_results}).

Assuming purely collisional excitation, and ignoring radiation
trapping, one can derive from solving the statistical equilibrium
equation that \citep{Tielens2005}
\begin{equation}
\frac{N_l}{N_u} = \frac{g_l}{g_u}
\left(\frac{n_{\rm{crit}}}{n_{\rm{H_2}}}+1\right)\,
\exp(h\nu/kT_{\rm{kin}})\,,
\label{Eqcol}
\end{equation}
where $N_u$, $N_l$, $g_u$, $g_l$ are the upper and lower state column
densities and degeneracies, $T_{\rm{kin}}$ the kinetic temperature,
$n_{\rm{crit}}$ the critical density, and $n_{\rm{H_2}}$ the hydrogen
density.  For a multi-level system, the critical density is given by
\citep{Tielens2005}
\begin{equation}
n_{\rm{crit}} = \frac{\sum_{l<u} A_{ul}}{\sum_{l \ne u}
  \gamma_{ul}}\,,
\end{equation}
where $A_{ul}$ represents the Einstein coefficient, and $\gamma_{ul}$
the collisional rate coefficients. Critical densities were calculated at
differing temperatures using the data available in the {\sc LAMDA}-database
\citep{Schoier2005A&A...432..369S} (see
Table~\ref{Table_results}).  Applying Eq.~(\ref{Eqcol}) together with
the column densities from Table~\ref{Table_results}, the minimum
density requirements for $n_{\rm{H_2}}$ to achieve the observed
number-density ratios are tabulated in Table~\ref{Table_NH2}. Note
that the listed values are at $T_{\rm{kin}}\,=\,300$\,K, and that for
lower temperatures the critical density increases.

\begin{table*}
\begin{minipage}[t]{\textwidth}
\caption{Velocity-integrated intensities (in K\,km/s), upper state
  column densities in cm$^{-2}$ calculated using Eq.~(\ref{EqNup}) and
  critical densites in cm$^{-3}$ for the detected multiple-transitions
  `parent' molecules used in the excitation analysis. In a few cases,
  extra integrated-intensity values (listed in italics) as found in
  literature are added to perform the excitation analysis. Literature
  references are given in the footnote.}\label{Table_results}
  \centering \setlength{\tabcolsep}{1mm}
  \renewcommand{\footnoterule}{} \renewcommand{\footnotesep}{-3ex}
\vspace*{-2ex}
\begin{tabular}{lcccccccccc}
\hline \hline
&  & SiO (5-4) & SiO (6-5) & SiO (7-6) & SiO (8-7) & HCN(3-2) &
HCN(4-3) & CS(5-4) & CS(6-5) & CS(7-6) \\
\hline
& $\int T_{mb}\,dv$ & 9.2 & 11.1 & 10.4 & & 9.9 & 8.9 & 2.4 &
[0.8]$^{\dagger}$ & 1.3 \\
\raisebox{1.5ex}[0pt]{WX Psc} & $N_u$  &  $7.96 \times 10^{12}$ & $7.91 \times 10^{12}$ &
$5.15 \times 10^{12}$ & & $7.96 \times 10^{12}$ & $4.22 \times
10^{12}$ & $ 2.53 \times 10^{12}$ & $1.08 \times 10^{12}$ & $1.41
\times 10^{12}$ \\
\hline
& $\int T_{mb}\,dv$ & 7.1 & 8.9 & 5.2 & & 21.7 & 12.2 &
\emph{8.5}$^a$ & 2.3 & 4.6 \\
\raisebox{1.5ex}[0pt]{W Aql} & $N_u$ & $6.19 \times 10^{12}$ & $6.36 \times 10^{12}$ & $2.57 \times 10^{12}$
& & $1.74  \times 10^{13}$ & $5.81 \times 10^{12}$ & $8.95 \times
10^{12}$ & $3.03 \times 10^{12}$ & $5.00  \times 10^{12}$ \\
\hline
& $\int T_{mb}\,dv$ & & & 19.62 &
\emph{20.4}$^b$ &
\emph{139.76}$^c$ & 60.69 & \emph{32.34}$^c$
& 20.37 & 35.24 \\
\raisebox{1.5ex}[0pt]{II Lup} & $N_u$ & & & $6.96 \times 10^{12}$ &
$8.74 \times 10^{12}$ & $1.24 \times 10^{14}$ & $2.88 \times 10^{13}$
& $6.85 \times 10^{13}$ & $2.61 \times 10^{13}$ & $3.83 \times
10^{13}$ \\
\hline
& $\int T_{mb}\,dv$ & 3.8 & 5.1 & &
\emph{8.4}$^d$ & 25.7 & \emph{31.9}$^d$ & 7.4
& & \\
\raisebox{1.5ex}[0pt]{V Cyg} & $N_u$ & $3.33 \times 10^{12}$ & $3.51
\times 10^{12}$ & & $5.28 \times 10^{12}$ & $2.06 \times 10^{13}$ &
$1.51 \times 10^{13}$ & $1.41 \times 10^{13}$ & & \\
\hline
\hline
  at 40\,K &  $n_{\rm{crit}}$ & $3.06 \times 10^{6}$ & $5.26 \times 10^{6}$ & $8.41
\times 10^{6}$ & $1.27 \times 10^{7}$ & $5.83 \times 10^{6}$ & $1.32
\times 10^{7}$ & $1.85 \times 10^{6}$ & $3.22 \times 10^{6}$ & $5.09
\times 10^{6}$ \\
 at 100\,K &  $n_{\rm{crit}}$ &$2.00 \times 10^{6}$ & $3.44 \times 10^{6}$ & $5.57
\times 10^{6}$ & $8.18 \times 10^{6}$ & $4.22 \times 10^{6}$ & $9.63
\times 10^{6}$ & $1.35 \times 10^{6}$ & $2.37 \times 10^{6}$ & $3.76
\times 10^{6}$ \\
  at 300\,K &  $n_{\rm{crit}}$ & $1.22 \times 10^{6}$ & $2.10 \times 10^{6}$ & $3.44
\times 10^{6}$ & $5.12 \times 10^{6}$ & $2.35 \times 10^{6}$ & $5.65
\times 10^{6}$ & $8.35 \times 10^{5}$ & $1.48 \times 10^{6}$ & $2.37
\times 10^{6}$ \\
\hline
\hline
\end{tabular}
$^a$ \citet{Bujarrabal1994A&A...285..247B}; $^b$
\citet{Schoier2006ApJ...649..965S}; $^c$
\citet{Woods2003A&A...402..617W}; $^d$
\citet{Bieging2000ApJ...543..897B}\hspace{\fill} \newline
$^{\dagger}$ line detected however with small S/N so that the  
integrated intensity is quite uncertain \hspace{\fill}
\end{minipage}
\end{table*}

\begin{figure*}
\begin{center}
\subfigure{\includegraphics[height=.45\textwidth,angle=90]{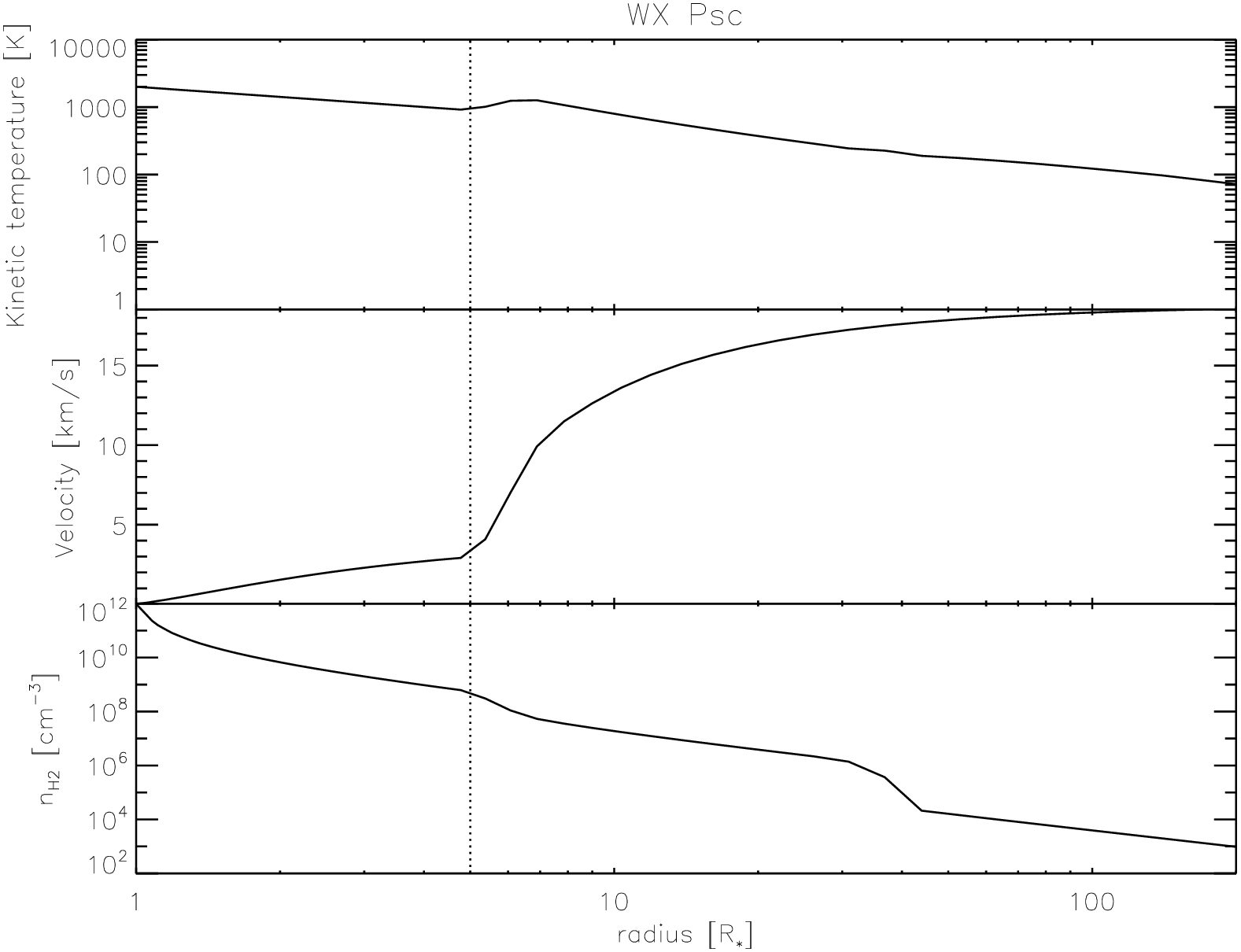}} \hspace{0.5cm}
\subfigure{\includegraphics[height=.45\textwidth,angle=90]{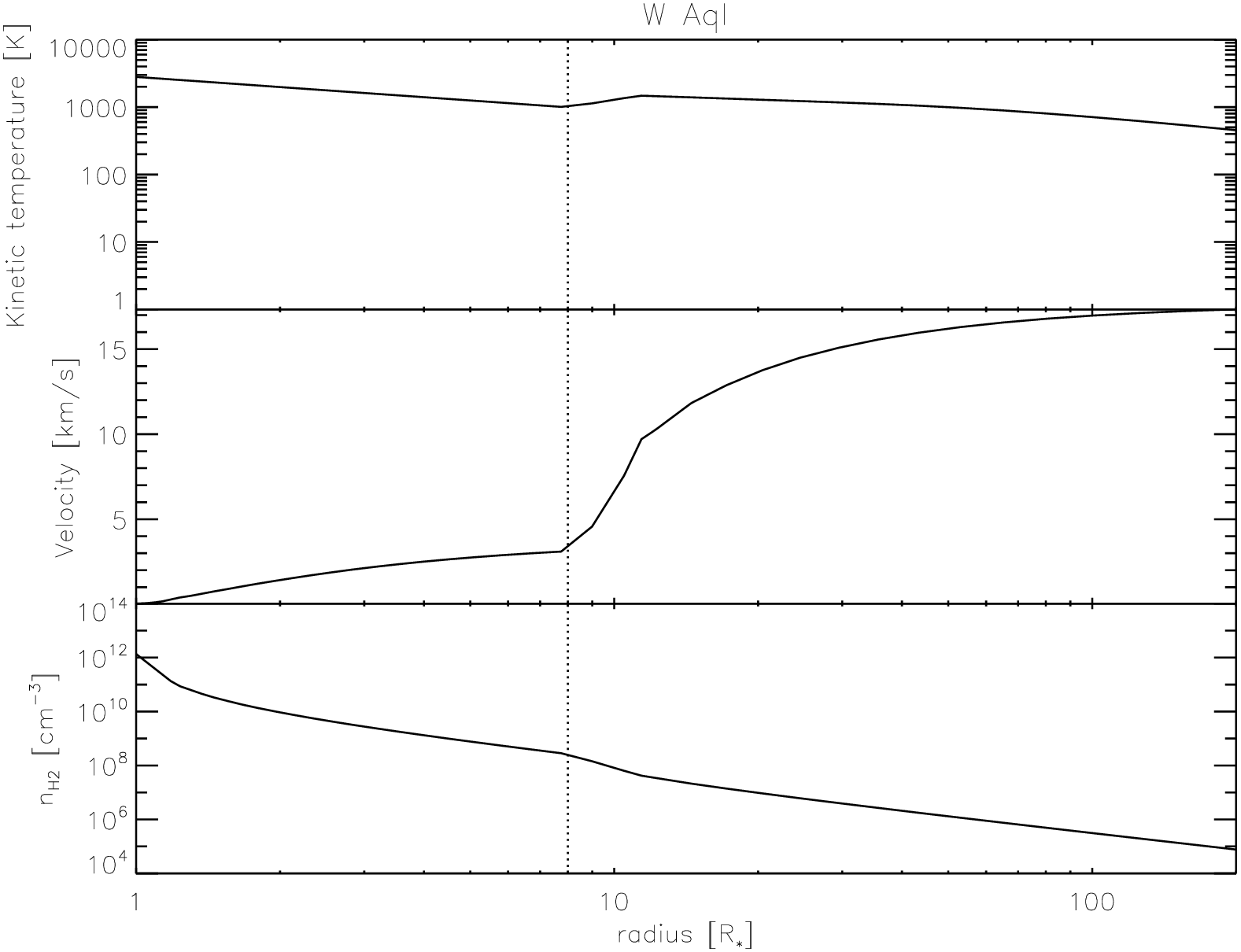}}
\subfigure{\includegraphics[height=.45\textwidth,angle=90]{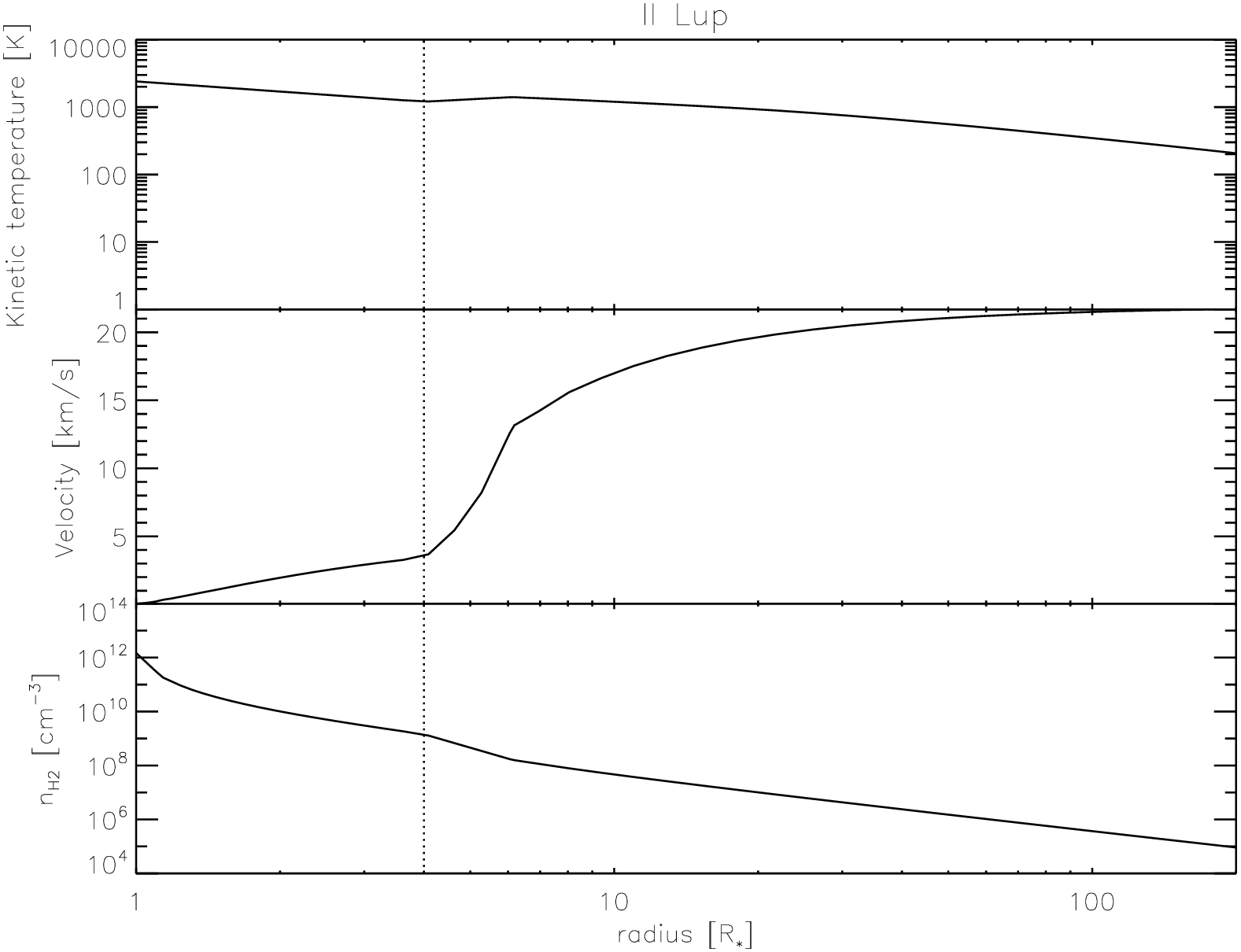}} \hspace{0.5cm}
\subfigure{\includegraphics[height=.45\textwidth,angle=90]{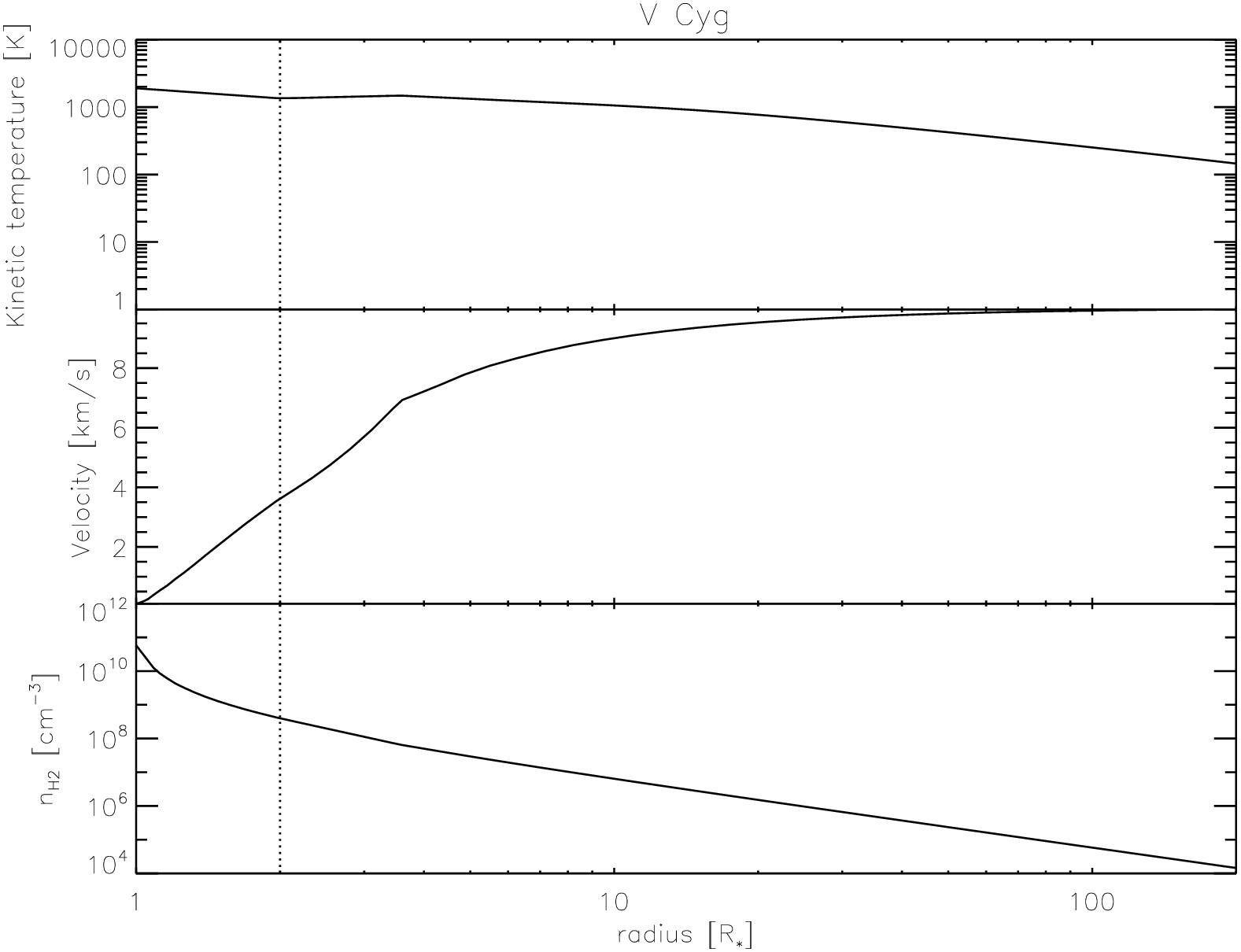}}
\caption{The structure of the CSE as derived using the GASTRoNOoM-code
is shown for the four studied targets \object{WX Psc}, \object{W Aql},
\object{II Lup}, and \object{V Cyg}. Since the focus is on the inner
and intermediate regions, only the first 200\,\Rstar\ are
displayed. \emph{Upper panel:} Estimated temperature profile,
\emph{middle panel:} estimated gas velocity structure, and
\emph{bottom panel:} estimated hydrogen density $n_{\rm{H_2}}$. The
dashed line indicates the dust condensation radius,
$R_{\rm{inner}}$. The variable mass-loss rate of \object{WX Psc},
mostly noticeable in the hydrogen density, is discussed in
\citet{Decin2007}.}
\label{Figstructure}
\end{center}
\end{figure*}

The gas temperature, density and velocity structure was calculated in
a self-consistent way using the GASTRoNOoM-code
\citep{Decin2006A&A...456..549D} to determine where in the envelope
the gas density falls below the density requirements given in
Table~\ref{Table_NH2} (see Fig.~\ref{Figstructure}). The assumed
stellar parameters are listed in Table~\ref{stellar_parameters}, and
the derived (maximum) radius of the emitting region is listed in
Table~\ref{Table_NH2}. If collisional excitation is assumed, it
appears from Table~\ref{Table_NH2} that the `parent' molecules SiO,
HCN and CS are excited in the inner ($\la 20$\,R$_*$) and intermediate
($\la 70$\,R$_*$ ) regions of the circumstellar envelope and trace
regions after the dust condensation zone where they have been injected
from deeper layers.

\begin{table}
\caption{For each target, the first row lists the minimum number
density for $n_{\rm{H_2}}$ in cm$^{-3}$ as derived using
Eq.~(\ref{Eqcol}) at $T_{\rm{kin}}\,=\,300$\,K and the second row
gives the corresponding maximum radius for the emission regions
obtained using the GASTRoNOoM-code. The third row gives the maximum
radius for the emitting region calculated using Eq.~(\ref{Eqrad}).}
  \label{Table_NH2} 
\centering 
\vspace*{-2ex}
\begin{tabular}{llccc}
\hline
\hline
 & & SiO & HCN & CS \\
\hline
 & $n_{\rm{H_2}}$ & $ 5.0 \times 10^6$ & $ 4.4 \times
10^6$ & $9.0 \times 10^5$ \\
WX Psc & R [Eq.~(\ref{Eqcol})] & $ 17$\,R$_*$ &
$ 19$\,R$_*$ & $ 33$\,R$_*$ \\
 & R [Eq.~(\ref{Eqrad})] & $ 4.5$\,R$_*$ &
$ 6$\,R$_*$ & $ 7.5$\,R$_*$\\
\hline
 & $n_{\rm{H_2}}$ & $ 2.1 \times 10^6$ & $ 2.1 \times
10^6$ & $ 6.34 \times 10^5$ \\
W Aql & R [Eq.~(\ref{Eqcol})] & $ 40$\,R$_*$ & $ 40$\,R$_*$ &
$ 70$\,R$_*$ \\
 & R [Eq.~(\ref{Eqrad})] & $ 9$\,R$_*$ & $ 10$\,R$_*$ & $ 11$\,R$_*$\\
\hline
  & $n_{\rm{H_2}}$ & $ 2.7 \times 10^7$ & $ 1.3  \times
10^6$ & $ 7.6 \times 10^5$ \\
II Lup & R [Eq.~(\ref{Eqcol})] & $ 13$\,R$_*$ & $ 53$\,R$_*$ &
$ 70$\,R$_*$ \\
 & R [Eq.~(\ref{Eqrad})] & $ 2.5$\,R$_*$ & $ 13$\,R$_*$ & $
10$\,R$_*$\\  
\hline
 & $n_{\rm{H_2}}$ & $ 2.8 \times 10^7$ & $ 8.7 \times 10^6$ & \\
V Cyg & R [Eq.~(\ref{Eqcol})] & $ 5$\,R$_*$ & $ 9$\,R$_*$ &  \\
 & R [Eq.~(\ref{Eqrad})] &  & $ 4$\,R$_*$ & \\
\hline
\hline
\end{tabular}
\end{table}

\begin{table}
\begin{minipage}[t]{\columnwidth}
\caption{Stellar parameters used as input for the GASTRoNOoM-code. The
  terminal velocity, $v_{\infty}$, is derived from the CO lines; the
  stellar radius from the stellar luminosity and temperature. The
  envelope density falls off as $\sim r^{-2}$. Literature references
  are given in the footnote.}
\label{stellar_parameters}
\centering
\vspace*{-2ex}
\begin{tabular}{lcccc}
\hline
\hline
& WX Psc & W Aql & II Lup & V Cyg \\
\hline 
T$_*$ [K] & 2000$^a$ & 2800 & 2400$^f$ & 1900$^f$ \\
R$_*$ [$10^{13}$\,cm] & 5.5 & 2.4 & 3.8 & 5.1 \\
L$_*$ [$10^3$\,L$_{\odot}$] & 10$^a$ & 6.8$^b$ & 8.8$^f$ & 6.3$^f$\\
{[CO]}/[H$_2$] [$10^{-4}$] & 3$^c$ & 6$^c$ & 8$^c$ & 8$^c$\\
distance [pc] & 833$^a$ & 230$^d$ & 500$^f$ & 310$^f$ \\
R$_{\rm{inner}}$ [R$_*$] & 5$^a$ & 8$^e$ & 4$^f$ & 2$^f$ \\ 
$v_{\infty}$ [km/s] & 18 & 17.5 & 21 & 10.5 \\
$\dot{M}$ [$10^{-6}$\,M$_{\odot}$\,yr$^{-1}$] & 6$^{a, \dagger}$ & 2.5$^b$ & 9$^f$ & 1.2$^g$\\
\hline
\hline
\end{tabular}
\phantom{a}\\
$^a$ \citet{Decin2007}; $^b$ \citet{Ramstedt2006A&A...454L.103R}; $^c$
\citet{Knapp1985ApJ...292..640K}; $^d$
\citet{Bieging2000ApJ...543..897B}; $^e$ \citet{Danchi1994AJ....107.1469D};
$^f$ \citet{Schoier2006ApJ...649..965S}; $^g$
\citet{schoier2001A&A...368..969S} \hspace{\fill} \newline
$^{\dagger}$ refers to the inner region (A) in \citet{Decin2007}
\hspace{\fill}   
\end{minipage}
\end{table}

It is also of interest to consider the case where collisional
excitation is ignored, and the molecules are excited by infrared
radiation from the star. One can derive that
\citep{Tielens2005}
\begin{equation}
\frac{N_l}{N_u} = \frac{g_l}{g_u} \frac{\exp(h \nu /k T_*) -1}{W} +
1\,,
\label{Eqrad}
\end{equation}
where $W$ is the geometrical dilution, being $(R_*/2R)^2$ when
R$\gg$R$_*$.  As seen from Table~\ref{Table_NH2}, radiative excitation
constrains the emitting zone for SiO being $\la$9\,R$_*$, for HCN
$\la$ 13\,R$_*$, and for CS $\la$11\,R$_*$. In general, these
regions are smaller than derived in case of collisional excitation,
indicating that the radiation field of the star does not have enough
energy to sustain the excitation.  An analogous expression as for
Eq.~(\ref{Eqrad}) can be derived for a radiation field characterized
by a dust temperature $T_d$ at the condensation radius
$R_{\rm{inner}}$, with the dilution factor then being
$(R_{\rm{inner}}/2R)^2$. Using the dust condensation radii listed in
Table~\ref{stellar_parameters} and a dust temperature of 800\,K the
derived emitting regions are a factor 3.2 larger for \object{WX Psc},
a factor 4.3 for \object{W Aql}, a factor 4 for \object{II~Lup}, and a
factor 2 for \object{V Cyg}, resulting in similar radii as in case of
collisional excitation.

Although the numbers in Table~\ref{Table_NH2} can only be used as
rough guidelines, they suggest in both cases a sequence in the
excitation pattern, SiO being the species emitting the closest to the
star, followed by HCN and CS.


\section{Conclusions} \label{Conclusions}

From the above analysis, one can draw the following conclusions:

\vspace{-1.5ex}
\begin{enumerate}
\item
The observations reported in this letter confirm the status of
`parent' molecules for CO, SiO, HCN, and CS in AGB stars, whose
observed molecular lines form close to the star in support of the
theoretical predictions of \citet{Cherchneff2006A&A...456.1001C}. The excitation
analysis suggests that SiO is emitted closest to the star, followed by
HCN and CS. 
\item High-excitation lines of SiS are detected in all stars, implying
that SiS too forms close to the star, whatever the stage of stellar
evolution. However, a thorough line analysis is necessary to prove or
disprove that SiS is more abundant in C stars than in O-rich Miras, as
predicted by \citet{Cherchneff2006A&A...456.1001C}.
\item SO appears to be typical of O-rich AGBs only. With only two
excitation lines detected, no information can be drawn on the locus of
SO formation. However, the SO($10_{11}$-$10_{10}$) line has an
estimated velocity of only 10\,km/s in \object{WX Psc} when the wind
terminal velocity for this object is 19.2\,km/s. This fact suggests
that the line excitation occurs close to star. In any case, the inner
chemistry of O-rich AGB envelopes appears to be as rich, if not
richer, than that of C-rich stars.

\end{enumerate}

A detailed line analysis of the
present data coupled to further observational campaigns with JCMT and
APEX are planned to corroborate these results.

\begin{acknowledgements}
   LD and SD acknowledge financial
  support from the Fund for Scientific Research - Flanders (Belgium),  IC acknowledges support from the Swiss National Funds for Science through
  a Marie-Heim-V{\"o}gtlin Fellowship, and 
  SH acknowledges financial support from the Interuniversity
  Attraction Pole of the Belgian Federal Science Policy P5/36. We
  thank Remo Tilanus (JCMT) for his support during the
  observations and reduction of the data.
\end{acknowledgements}

\vspace*{-.5cm}

\end{document}